\documentclass[amsmath,byrevtex,titlepage,
preprint,
aps, superscriptaddress,
nofootinbib ]{revtex4-1}
\usepackage{graphicx}% Include figure files
\usepackage{dcolumn}% Align table columns on decimal point
\usepackage{bm}% bold math
\usepackage{amsmath}
\usepackage{amsfonts}
\usepackage{amssymb}
\usepackage{hyperref}

\newcommand{\eq}[1]{eq.~(\ref{#1})}

\newcommand{\beq}{\begin{equation}}
\newcommand{\eeq}{\end{equation}}

\newcommand{\la}[1]{\label{#1}}
\newcommand{\bea}{\begin{eqnarray}}
\newcommand{\eea}{\end{eqnarray}}
\newcommand{\ba}{\begin{array}}
\newcommand{\ea}{\end{array}}

\begin{document}
%\rightline{\textsl{\date{\today}}} \vspace{0.5cm} %
\title{ Narrow Nucleon-$\psi(2S)$ Bound State and LHCb Pentaquarks}
\author{Michael I.~Eides}
\affiliation{Department of Physics and Astronomy, University of Kentucky, Lexington, KY 40506, USA}
\affiliation{Petersburg Nuclear Physics Institute, Gatchina, 188300, St.Petersburg, Russia}
\author{Victor Yu.~Petrov}
\affiliation{Petersburg Nuclear Physics Institute, Gatchina, 188300, St.Petersburg, Russia}
\author{Maxim V.~Polyakov}
\affiliation{Petersburg Nuclear Physics Institute, Gatchina, 188300, St.Petersburg, Russia}
\affiliation{Institut f\"ur Theoretische Physik II, Ruhr-Universit\"at Bochum, D - 44780 Bochum,
Germany}

\begin{abstract}

We interpret the newly discovered pentaquark $P_c(4450)$ as a bound state of charmonium $\psi(2S)$ and the nucleon. The binding potential is due to charmonium-nucleon interaction that in the heavy quark approximation is proportional to the product of the charmonium chromoelectric polarizability and the nucleon energy-momentum distribution. We use the large $N_c$ expansion to estimate the quarkonium polarizability and calculate the nucleon properties in the framework of the mean-field picture of light baryons. Two almost degenerate states $J^P=(1/2)^-$ and $J^P=(3/2)^-$ are predicted at the position of the $P_c(4450)$ pentaquark.  We find that the nucleon-$\psi(2S)$ bound state has a naturally narrow width in the range of tens of MeV. The unitary multiplet partners of the $P_c(4450)$ pentaquark and the generalization to $b \bar b$-nucleon pentaquark bound states are discussed.

\end{abstract}

\maketitle

\section{Introduction}

The discovery of new pentaquark states by the  LHCb collaboration \cite{LHCb} poses the problem of their internal structure. A few interesting ideas were already proposed: the pentaquark as a loosely bound state  of charmed baryon and meson \cite{molecula}, the pentaquark as a bound state of light and heavy diquarks with a $c$-quark \cite{diquarks}, and even the pentaquark as a bound state of  states with open color \cite{Morozov}. It was also suggested in \cite{cusp} that the structures found by LHCb can be interpreted as threshold cusp effects.

In this letter we explore another option: pentaquark as a bound state of a charmonium state and the nucleon. A heavy quark-antiquark bound state is a small (compared to the size of a nucleon) heavy neutral object. Its interaction with a nucleon is relatively weak even when the distance between quarkonium and nucleon is small. Quarkonium can easily  penetrate the nucleon and form a true pentaquark state. In this state the distances between the three quarks of the nucleon and the compact heavy meson are all of the same order.

Nonrelativistic multipole expansion is a natural framework for discussion of strong interactions of a heavy quarkonium \cite{gott1978}. This interaction is dominated by virtual emission of two chromoelectric dipole gluons in a color singlet state. Then the effective interaction potential between the heavy quarkonium and the nucleon is proportional to the product of the meson chromoelectric polarizability and the local gluon energy-momentum density inside the nucleon \cite{Voloshin}.

Quarkonium chromoelectric polarizability was widely discussed for many years. It can be reliably calculated theoretically in the heavy quark mass limit when the quarkonium can be considered as a Coulomb system. This calculation \cite{Peskin} for any quarkonium energy level was  done in the large $N_c$ approximation. Nondiagonal (transitional) polarizabilities also can be calculated in this approach. It is debatable how close the real heavy quark systems ($c\bar{c}$ or $b\bar{b}$ quarkonia) are to the pure Coulomb system. Phenomenological values of the transitional polarizabilities can be extracted, e.g.,  from  the experimental data on the $\psi'\to J/\psi\pi\pi$ decays \cite{Voloshin}. There is at least a qualitative agreement between the Coulombic and phenomenological values of nondiagonal polarizabilities.

A model is needed to estimate the gluon energy-momentum density inside a nucleon.  The simplest but not too accurate estimate is provided by the Skyrme soliton model \cite{Witten}. The QCD inspired Chiral Quark-Soliton Model ($\chi QSM$) \cite{we} was very successful in describing virtually all low-energy physics of interacting nucleons and pseudoscalar mesons \cite{Diakonov:2000pa}. It arises in QCD in the large $N_c$ limit and unambiguously leads to the mean-field picture of baryons \cite{mfb}. We calculate the gluon energy-momentum density inside a nucleon in this  mean-field framework.

The effective quarkonium interaction with light hadrons described above turns out to be attractive. It was used to discuss the possibility of quarkonium  bound states in light nuclear matter \cite{gg}. It was also applied to interpretation of exotic mesons with hidden charm \cite{hadrocharmonium}. A tentative interpretation of the LHCb pentaquarks as bound states of $J/\psi$  and the nucleon resonances $N(1450)$ and $N(1520)$ was suggested in \cite{kubvol2015}.

The strength of the quarkonium-nucleon interaction is determined by the quarkonium polarizability. We will see below that the interaction seems to be not strong enough to bind together $J/\psi$ and an individual nucleon.  Coulombic chromoelectric polarizability increases like cube of the quarkonium radius. One can hope that the fast growth of polarizability with radius of the heavy quark-antiquark bound state holds even for non Coulombic systems. As a result interaction of a nucleon with excited quarkonia is much stronger than interaction interaction with $J/\psi$,  and bound nucleon-excited quarkonia states should exist.

We will show below that the strength of the attraction potential between the soliton and the excited $\psi(2S)$  state is about a few hundreds MeV and the size of this potential is about 1 fm. We interpret the bound state in this potential as the $P_c(4450)$ pentaquark discovered by the LHCb collaboration. The pentaquark in this picture has a rather small width about a few tens MeV.

\section{Interaction of heavy quarkonium with nucleon}

It was understood long time ago that the interaction of heavy quarkonium with light hadrons is due to the soft gluon fields and can be described in the framework of multipole expansion \cite{volosh1982}. The
role of a small parameter in this expansion plays the ratio of quarkonium size over the effective gluon
wavelength. The leading term in this expansion is due to two dipole gluons and can be parameterized in
terms of chromoelectric polarizability $\alpha$. The effective dipole  Lagrangian has the form \cite{volosh1982}

\beq \label{dipham}
L_{eff}=\frac{\alpha}{2}\ \bm E\cdot \bm E,
\eeq

\noindent
where $\bm E$ is the chromoelectric gluon field (with the coupling constant absorbed), and $\alpha$ is the chromoelectric polarizability.

As we already mentioned the chromoelectric polarizabilities of charmonium states are not known now, except in the case of very heavy quarks. For such quarks one can consider quarkonium as a Coulombic system and polarizability admits perturbative calculation in the framework of the $1/N_c$ expansion \cite{Peskin}. After calculations we obtain polarizability for an arbitrary quarkonium $nS$ energy level

\beq \label{alphacoulomb}
\alpha(nS)=\frac{16\pi n^2}{3g^2N_c^2}c_na_0^3,
\eeq

\noindent
where $c_1=7/4$, $c_2=251/8$, $c_n(n\geq 3)=(5/16)n^2(7n^2-3)$, $a_0=16\pi/(g^2N_cm_q)$ is the Bohr radius of nonrelativistic quarkonium, and $g$ is the coupling constant normalized at the size of quarkonium.  The nondiagonal $(2S\to1S)$ chromoelectric polarizability turns out to be

\beq \label{alphacoulomb1}
\alpha(2S\to1S)=-\frac{51200\sqrt{2}\pi}{1287g^2N_c^2 }a_0^3.
\eeq

\noindent
Other transitional polarizabilities can be calculated in the same way.

We use the Coulombic values for polarizabilities as an order of magnitude estimates of their scale and characteristic features but we will not rely heavily on their numerical values. For the numerical estimates we assume that $J/\psi$ and $\psi'$ may be considered as nonrelativistic Coulomb bound states. Fitting the energy levels we extract the Bohr radius and obtain polarizabilities\footnote{ The result may vary slightly depending on how one treats large $N_c$ limit.

%\bf I obtain
%
%\beq
%\alpha(1S)\approx 0.3~ {\rm GeV}^{-3}, \qquad \alpha(2S)\approx 20~ {\rm
%GeV}^{-3},\qquad \alpha(2S\to 1S)\approx -1.7~{\rm GeV}^{-3}.
%\eeq
%
%if I use $N_c=\infty$ in the expression for the Coulomb coupling constant.
%
%If I use $SU(3)$ expression for the Coulomb coupling constant I obtain
%
%\beq
%\alpha(1S)\approx 0.14~ {\rm GeV}^{-3}, \qquad \alpha(2S)\approx 10~ {\rm
%GeV}^{-3},\qquad \alpha(2S\to 1S)\approx -0.9~{\rm GeV}^{-3}.
%\eeq
}

\beq \label{alphanum}
\alpha(1S)\approx 0.2~ {\rm GeV}^{-3}, \qquad \alpha(2S)\approx 12~ {\rm
GeV}^{-3},\qquad \alpha(2S\to 1S)\approx -0.6~{\rm GeV}^{-3}.
\eeq

\noindent
Transitional polarizability $|\alpha(2S\to 1S)|\approx 2~{\rm GeV}^{-3}$ was extracted from the phenomenological analysis of the $\psi'\to J/\psi\pi\pi$ transitions \cite{Voloshin}. There is a rather significant discrepancy between the perturbative result above and this value. It could be explained by the noncoulombic nature of quarkonium. We expect that calculations with a more realistic potential would lead to a better agreement with the phenomenological value of polarizability.

The chromoelectric field squared in the Lagrangian in \eq{dipham} can be easily connected with the gluon part of the QCD energy-momentum tensor $T_{00}^G$ and, via the conformal anomaly, with the trace of the full energy-momentum tensor ${T^\mu}_\mu$\footnote{We ignore the contribution of the light quarks mass term. Simple estimates show that this term shifts the mass of the pentaquarks by
only about 10~MeV upwards and hence can be safely neglected for all practical purposes.}

\[ \label{E2}
\bm E^2
=\frac{\bm E^2-\bm H^2}{2}+\frac{\bm E^2+\bm H^2}{2}=g^2\left(\frac{8\pi^2}{bg^2_s}{T^\mu}_\mu + T^G_{00}\right).
\]

\noindent
Here $b=({11}/{3}) N_c-(2/3) N_f$ is the leading coefficient of the Gell-Mann-Low
function, $g_s$ is the strong coupling constant at a low normalization point. Notice that due to running of the coupling constant in QCD $g\neq g_s$.  The coupling constant  $g$ is defined at the scale of the quarkonium radius, while $g_s$ is defined at the scale of the nucleon radius. It seems that we can safely ignore this distinction in the case of charmonium but it could become important for bottomonium.

Now we are ready to adjust the effective Lagrangian in \eq{dipham} for analysis  of the quarkonium interaction with a light hadron. To this end we average the operator in \eq{dipham} over the hadron state and obtain

\beq \label{HeffEMT}
{\cal L}_{eff}=\frac{\alpha}{2}g^2\left(\frac{8\pi^2}{bg_s^2}{T^\mu}_\mu + T^G_{00}\right)=\frac{\alpha}{2}g^2\left(\frac{8\pi^2}{bg^2_s }{T^\mu}_\mu +
\xi  T_{00}\right),
\eeq

\noindent
where ${T^\mu}_\mu$ and $T_{00}$ are now expectation values of the respective operators in the light hadron state. At the last step we also introduced a new parameter $\xi$ that describes the fraction of the nucleon energy carried by the  gluons at a low normalization point, $T^G_{00}=\xi  T_{00}$.

We analyze the quarkonium-nucleon interaction with the help of the effective interaction Lagrangian in \eq{HeffEMT} using the $\chi QSM$ model of the nucleon and the estimates of the chromoelectric polarizabilities above. Both the heavy quarkonium and the nucleon in the large $N_c$ limit are nonrelativistic. In these conditions the interaction Lagrangian in \eq{HeffEMT} describes a static interaction. The respective nonrelativistic potential can be written in terms of the local energy density $\rho_E(\bm x)$ and pressure  $p(\bm x)$ \cite{forces}

\beq \label{effpot}
V(\bm{x}) = -\alpha\frac{4\pi^2}{b }\left(\frac{g^2}{g_s^2}\right)\left[ \rho_E(\bm x) \left(1+\xi\frac{bg^2_s}{8\pi^2}\right) -
3 p(\bm x)\right].
\eeq

\noindent
This effective potential has a simple interpretation. A point-like quarkonium serves as a tool that scans the local energy density and local pressure inside the nucleon. It could happen that the size of quarkonium is not small enough in comparison with the size of the nucleon. In such case we will need to consider higher order terms in the QCD multipole expansion in order to improve description of the quarkonium-nucleon interaction.

The overall normalization of the effective potential

\beq \label{intV}
\int d^3x V(\bm{x}) = -\alpha
\frac{4\pi^2}{b }\left(\frac{g^2}{g_s^2}\right){ M}_N \left(1+\xi\frac{bg_s^2}{8\pi^2}\right)
\eeq

\noindent
is determined by the total energy of the nucleon $\int d^3 x \rho_E (\bm x)=M_N$ and the
stability condition $\int d^3x p(\bm x)=0$. The factor $\nu= 1+\xi({bg_s^2}/{8\pi^2})$ is model dependent. An estimate of this factor for the pion in \cite{Shifman} produced $\nu\sim 1.45-1.6$.  In the theory of instanton vacuum and the $\chi QSM$ model the strong coupling constant freezes at the size of the nucleon with the value about $\alpha_s=g_s^2/4\pi\sim0.5$. Using this coupling constant we obtain $\nu \sim 1.5$ for the nucleon, that is close to the pion result in \cite{Shifman}.

The large distance behavior of the potential in \eq{effpot} is determined by the leading term in the asymptotic expansion of the pion mean field in the nucleon. This term can be calculated in a model-independent way and in the chiral limit ($m_\pi=0$) the potential at large distances has the form

\beq \label{effpotasym}
V(\bm{x}) \sim -\alpha\frac{27 (1+\nu)}{16 b } \frac{g_A^2}{F_\pi^2 |\bm x|^6}.
\eeq

\noindent
Here $g_A\approx 1.23$ is the nucleon axial charge, and $F_\pi\approx 93$~MeV is the pion decay constant.

The local energy density $\rho_E(\bm x)$ and pressure $p(\bm x)$ were computed in the $\chi$QSM in \cite{maxim}. Calculations involved the exact quark levels in the pion mean field (including the Dirac sea) and solution of the self-consistent equations of motion for the mean field. In this approach  the normalization condition for the potential in \eq{intV} is satisfied automatically since the normalization condition for the energy density and the stability condition for the pressure hold in the self-consistent calculation due to the equations of motion.

%The form of the resulting potential for the nucleon-$\psi(2S)$ potential  with chromo-electric polarizability of $\alpha=$

 %is plotted in the Fig.1. We see that the potential is rather deep and can bind a quarkonium.

%\begin{figure} %\includegraphics{potemtial.eps} %\caption{Nucleon-$\psi(2S)$ potential (in MeV) as a function of distance (in fm). } %\la{fig1} %\end{figure}

%\begin{figure} %\vspace*{-0.4cm}
%\centerline{\epsfverbosetrue\epsfxsize=9.5cm\epsfysize=7.5cm\epsfbox{potemtial.eps}}
%\includegraphics{as1}% Here is how to import EPS art %\caption{Nucleon-$\psi(2S)$ potential (in MeV) as a function of distance (in fm) for the case of $\alpha(2S)=106$~GeV$^{-3}$ as predicted %by the heavy quark mass-- and large $N_c$-- limits. The energy density and distribution of the pressure are taken from Ref.~\cite{maxim}. } %\label{fig1} \vspace{-0.3cm} %\end{figure}

\section{Mass of nucleon-$\psi(2S)$ bound state}

We have found the nonrelativistic quarkonium-nucleon interaction potential in terms of the local nucleon energy density and pressure and chromoelectric polarizability $\alpha$. The form of this potential in \eq{effpot} is determined by results of the self-consistent mean-field calculation in \cite{maxim}, its overall strength is fixed by the values of the chromoelectric polarizabilities of quarkonia. Notice that this potential is universal, interaction of any quarkonium state with the nucleon is described by a potential with one and the same functional form, only the overall normalization depends on the quarkonium energy levels. Even the potentials that describe nondiagonal transitions between the quarkonium  states have the same form. The quarkonium-nucleon potentials for the two lowest charmonium states have the form

\beq \label{channelpotential}
V_{22}(r)\equiv V(r), \qquad V_{11}(r)=\frac{\alpha(1S)}{\alpha(2S)}V(r), \qquad
V_{12}(r)=\frac{\alpha(2S\to 1S)}{\alpha(2S)}V(r),
\eeq

\noindent
where $V(r)$ is the potential in \eq{effpot} with $\alpha=\alpha(2S)$, and other potentials are scaled by the ratios of the respective chromoelectric polarizabilities. With the values of polarizabilities from \eq{alphanum}, the potentials $V_{11}(r)$, $V_{12}(r)$ are small in comparison with the potential $V(r)$. The potential $V_{12}(r)$ describes the transition $J/\psi\to\psi'$ off the nucleon.

Possible bound states in the channels  $J/\psi N$ and $\psi' N$ are solutions of the Schr\"odinger equation

\beq \label{Sch1}
\left(-\frac{\bm\nabla^2}{2\mu}+ V(r)-E\right)\Psi_b  = 0,
\eeq

\noindent
where $\mu$ is the reduced mass in the respective channel and the potential is defined in \eq{effpot}. Due to the poor knowledge of the chromoelectric polarizability $\alpha$ we can vary it in a relatively wide region.

Solving the eigenvalue problem in \eq{Sch1} we found that:

\begin{enumerate}

\item A bound state arises when the chromoelectric polarizability reaches the critical value  $\alpha = 5.6$~GeV$^{-3}$. Comparing this value with the Coulomb values in \eq{alphanum} we see that $J/\psi$ does not form a bound state with the nucleon. For the excited charmonia states  $\psi(2S)$, $\psi(3S)$, etc. the critical value of $\alpha$ is far below the expected chromoelectric polarizabilities of the excited charmonia. Therefore, they seem to form bound
    states with the  mean-field nucleon. Below we will consider the bound state(s) of $\psi(2S)$, higher excited charmonia will be considered elsewhere.

\item A bound state with the orbital     momentum $l=0$ and with the binding energy $E_b=-176$~MeV (corresponding to the position of     the $P_c^+(4450)$ pentaquark) is formed at $\alpha(2S)=17.2$~GeV$^{-3}$. There are no other bound
    states in this case.

\item A bound state with the orbital momentum $l=0$ and with the
    energy $E_b=-246$~MeV (corresponding to the position of the $P_c^+(4380)$ pentaquark) is formed
    at $\alpha=20.2~$GeV$^{-3}$. Again, there are no other bound states in this case. Hence, if try to interpret $P_c^+(4380)$ as a bound state with $E_b=-246$~MeV, there would be no place for heavier pentaquarks to be observed in the $J/\psi+N$ channel.

\item
At a bit larger value of polarizability $\alpha\approx 22.4$ a bound state with angular momentum $l=1$ arises for the first time. One could try to identify the light pentaquark with the $l=0$ bound state and the heavy pentaquark with the $l=1$ bound state. The quantum numbers of such pentaquarks would be $({3}/{2})^-$ and $({5}/{2})^+$, what fits the experimental data nicely. However, we consider this option to be absolutely excluded. The mass difference of the states with $l=1$ and $l=0$ is about 300~MeV, not the observed 70~MeV. This large mass difference is due a relatively small size (around $~0.8-0.9$ fm) of the nucleon. Respectively,  the nucleon moment of inertia is small, and the energy of its rotational excitations is about a few hundred MeV as it can be seen from $N-\Delta$ mass difference. Additionally, the scenario with two pentaquarks as the $l=0$ and $l=1$ bound states cannot explain the widths of the observed pentaquarks.

\end{enumerate}

We see that for reasonable values of the chromoelectric polarizability $\alpha(2S)$ the charmonium $\psi(2S)$ binds with the  mean-field nucleon. Notice, however, that for a given value of $\alpha(2S)$ only one bound level exists. It means that the picture we suggest here can describe only one of the LHCb pentaquarks. Experimentally $P_c(4380)$ has a rather large width $205\pm18\pm 86$~MeV, whereas the $P_c(4450)$ is rather narrow with the width $39\pm5\pm 19$~MeV. We will see in next section that the nucleon-$\psi(2S)$ bound state has a naturally narrow width about a dozen MeV. Therefore, the interpretation of the nucleon-$\psi(2S)$ bound state as the LHCb $P_c(4450)$ pentaquark seems to be more justified.

The nucleon-$\psi(2S)$ bound state is formed in the $S$-wave, hence its quantum numbers can be either $J^P=(1/2)^-$ or $J^P=(3/2)^-$. The hyperfine splitting between the color singlet states arises due to interference of the chromoelectric dipole $E1$ and the chromomagnetic quadrupole $M2$ transitions and can be described by an effective Hamiltonian

\beq
H_{eff}
=-\frac{\alpha}{4m_q}S_j\langle N|[E^a_i(D_iB_j)^a+(D_iB_j)^aE^a_i]N\rangle,
\eeq

\noindent
where $S_j$ is the quarkonium spin, $\alpha$ and $m_q$ are the same chromoelectric polarizability and the heavy quark mass as above, and only the nucleon matrix element of the product of chromoelectric and chromomagnetic fields requires calculation. We see that the quarkonium-nucleon spin-spin interaction is suppressed by the heavy quark mass $\sim 1/m_q$, therefore in the leading order of the heavy quarks expansion the $(1/2)^-$ and $(3/2)^-$  states are degenerate.  A semiquantitative estimate of this splitting produces a small value in the range of $5-10$~MeV. Therefore we predict that there are actually two almost degenerate pentaquark states with $J^P=(1/2)^-$ and $J^P=(3/2)^-$  at the position of the observed pentaquark at $M_{pJ/\psi}=4450$~MeV. It would be very interesting if the LHCb collaboration could check this hypothesis in their partial wave analysis.

\section{The partial width of the nucleon-$\psi(2S)$ bound state}

Let us calculate the partial decay width of the pentaquark to $J/\psi+N$. To this end
we consider $J/\psi$ scattering off the nucleon as a nonrelativistic two-channel problem

\beq
\begin{split} \label{2channel}
&\left(-\frac{\bm\nabla^2}{2\mu_1}+ V_{11}(r)-E\right)\Psi_1 + V_{12}(r)\Psi_2 =0,
\\
&\left(-\frac{\bm\nabla^2}{2\mu_2}+ V_{22}(r)-E+\Delta\right)\Psi_2 + V_{12}(r)\Psi_1 = 0.
\end{split}
\eeq

\noindent
Here $\mu_1$ and $\mu_2$ are the reduced masses of $J/\psi+N$ and $\psi'+N$ respectively,
$E$ is the energy in the center of mass frame ($E = \bm p^2/2\mu_1$, where $\bm{p}$ is the relative momentum), $\Delta=M_{\psi'}- M_{J/\psi}$, and the potentials $V_{11}(r)$, $V_{22}(r)$, $V_{\psi',J/\psi}$ are defined in \eq{channelpotential}.

Due to the non-zero transition potential $V_{\psi',J/\psi}$ the pentaquark arises as a resonance in the $J/\psi N$ scattering channel described by the standard Breit-Wigner formula. We will find the width of the resonance from this resonance scattering amplitude.

The transition potential $V_{\psi',J/\psi}$ is small and we solve the scattering problem in \eq{2channel} using perturbation theory. In the leading approximation the wave function $\Psi_1(\bm x)$ is just an incoming plane wave $e^{i{\bm q}\cdot {\bm x}}$ where ${\bm q}$ is the center-of-mass momentum before scattering. Due to the coupling $V_{\psi',J/\psi}$ between the channels this plane wave leaks in the second channel where it induces the wave function

\beq \label{leakkf}
\Psi_{2}({\bm x})=-\int d^3x^{\prime}G_2({\bm x},{\bm x'})V_{12}({\bm x}^{\prime})e^{i {\bm q}\cdot{\bm
x}^{\prime}}.
\eeq

\noindent
Here

\beq
G_2({\bm x},{\bm x'})=\left\langle {\bm x}\left|\frac{1}{-\frac
{\bm\nabla^{2}}{2\mu_2}-E+\Delta+V-i0}\right|\bm{x}^{\prime}\right\rangle
\eeq

\noindent
is the Green function of the Schr\"odinger equation for $\Psi_2(\bm x)$ (see \eq{Sch1}). Near the resonance it can be approximated by

\[
G_2({\bm x},{\bm x'}) = \frac{\psi_{R}({\bm x})\psi_{R}^*({\bm x'})}{E_R-E},
\]

\noindent
where $E_R$ is the resonance energy. The wave function  $\Psi_{2}({\bm x})$ in \eq{leakkf} in its turn generates correction to $\Psi_{1}({\bm x})$. This correction has the from (see the first line in \eq{2channel}) 

\beq \label{corr2}
\delta\Psi_1(x) = \int d^3 x' G_1({\bm x},{\bm x'})V_{12}({\bm x}')\psi^*_R({\bm x}') \frac{\int d^3 x''
V_{12}({\bm x}'')\psi_R({\bm x}'')e^{i {\bm q}\cdot{\bm x}^{\prime\prime}}}{E_R-E},
\eeq

\noindent
where $G_1(\bm x, \bm x^\prime)$ is the Green function of the free Schr\"odinger equation,

\beq G_1({\bm
x},{\bm x'})= 2\mu_1 \frac{e^{i q |{\bm x}-{\bm x}'|}}{4\pi|{\bm x}-{\bm x}'|}.
\eeq

\noindent
Here $\bm q$ is the
center of mass momentum corresponding to the resonance energy, $|\bm q|=q=\sqrt{2 \mu_1 E_R}$.

At  large $x\to\infty$ the wave function $\delta\Psi_1(x)$ is just an outgoing spherical wave. Then the wave function in the first channel at large $x$ is a superposition of the incoming plane wave and an outgoing spherical wave

\beq
\Psi_1({\bm x})+\delta\Psi_1(x) = e^{i{\bm
q}\cdot {\bm x}} + f(\theta) \frac{e^{iqr}}{r},
\eeq

\noindent
where $f(\theta)$ is the scattering amplitude  ($\theta$ is the scattering angle). The scattering amplitude as determined by the wave function \eq{corr2} has a standard Breit-Wigner resonance form

\beq
f(\theta)=-\frac{2l+1}{q}\frac{\Gamma/2}{E-E_R} P_l(\cos\theta),
\eeq

\noindent
where $\Gamma$ is the partial decay width of the resonance into the $N+J/\psi$ channel. Calculating the width we obtain

\beq
\Gamma
=\left(\frac{\alpha(2S\to1S)}{\alpha(2S)}\right)^2(4\mu_1 q)\left|\int_0^\infty dr r^2 R_l(r)V(r)
j_l(q r)\right|^2, \la{gamma}
\eeq

\noindent
where $R_l(r)$ is the radial wave function of the resonance normalized by the condition $\int drr^2R_L(r)=1$, and $j_l(z)$ is the spherical Bessel function.

We obtain the partial decay width $\Gamma(P_c(4450)\to N+J/\psi)=11.2$~MeV using the phenomenological value of polarizability $\alpha(2S\to 1S)=2$~GeV$^{-3}$ based on the analysis of the $\psi'\to J/\psi\pi\pi$ transitions in \cite{Voloshin}. We also made a rough estimate of the width of the decay $P_c\to J/\psi+N+\pi$, and it turns out to be even smaller than the partial width into the $J/\psi+ N$ channel. The decays of the pentaquark into (anti)charmed meson + charmed baryon should be strongly suppressed  in the scenario above, since decays of the pentaquark into open charm channels  require exchange of a heavy $D$-meson in the $t$-channel and appear to be small for this reason. Therefore the total width of the $P_c$ pentaquark in our picture is small -- in the range of tens MeV, in excellent agreement with the experimentally observed width $\Gamma_{\rm exp}=39\pm 5\pm 19$~MeV of the $P_c(4450)$ pentaquark.

\section{Conclusions and outlook}

We interpret the newly discovered pentaquark $P_c(4450)$ as a bound state of $\psi(2S)$ and the nucleon. The binding is due to chromoelectric interaction between a small quarkonium state and  the nucleon. The nucleon is described in the framework of the mean-field picture of light baryons in the $\chi QSM$ model. Let us mention that the $\Theta^+$ pentaquark \cite{dia1} and the charmed pentaquark \cite{dia2} were earlier predicted in the $\chi QSM$ model. However, the physical nature of those pentaquarks is completely different from the mechanism considered here. The two main ingredients of the present discussion, small size of quarkonium and quarkonium-nucleon interaction, played no role in those predictions.

We used the large-$N_c$ limit and the heavy quark mass approximation, when charmonium interacts with the local energy density and pressure of the nucleon. These nucleon characteristics  were calculated in the $\chi QSM$ model in \cite{maxim}. The strength of the charmonium interaction with the nucleon mean-field is determined by the charmonium chromoelectric polarizability $\alpha(2S)$.
The charmonium-nucleon bound state arises at reasonable values of $\alpha(2S)$, the chromoelectric polarizability can be adjusted is such way that the bound state mass coincides with the position of either $P_c(4380)$ or $P_c(4450)$. Let us emphasize that only one $\psi(2S)$-nucleon bound state arises in our approach. The possibility that the nucleon binds with higher excited states of charmonia ($\psi(3S)$, etc)  will be considered elsewhere.

We have demonstrated that the nucleon-$\psi(2S)$ bound state has a naturally narrow width in the range of tens MeV.  Therefore the wide $P_c(4380)$ pentaquark does not fit into our picture, it seems that it should be of other nature. We predict that the $P_c(4450)$ peak  consists of  two almost degenerate pentaquark states with $J^P=(1/2)^-$ and $J^P=(3/2)^-$. This is at variance with the most favorable quantum number of $J^P=(5/2)^+$ obtained for this pentaquark in the analysis of the LHCb collaboration \cite{LHCb}.

The possibility that $c\bar{c}$ resonances can bind with baryons would open ad rich world of pentaquarks. The presence of a compact weakly interacting particle inside the baryon does not change its properties in a significant way. This means that the pentaquark states should duplicate all already known baryon multiplets. For example, the pentaquark $P_c(4450)$ discovered by LHCb should be a member of a baryon octet.  Masses of other particles in this octet can be read off the table of baryons: we expect analogues of $N$, $\Sigma$, $\Xi$ and $\Lambda$. The next multiplet of pentaquarks is similar to the baryon decuplet and should consist of  pentaquarks with the properties similar to $\Delta,\Sigma,\Xi,\Omega$. This is also not the end of the story -- we see no reason why
$\psi(2S)$ cannot form a bound state with the Roper resonance or any other known baryon with positive or negative parity.

The other opportunity to proliferate the number  of pentaquark states even more is to consider possible bound states of baryons with other excited states of $c\bar{c}$ systems. It is also worth noticing that the spin-spin interaction between $c\bar{c}$-mesons and nucleons is very weak. This means that every pentaquark state should be accompanied by nearly degenerate states with different spins (but not with the same parity).

We were discussing here  $c\bar{c}$ systems but one can repeat all this for $b\bar{b}$ ones. Moreover, our considerations should become more reliable for systems with $b$-quarks as they are definitely closer to pure Coulomb systems. On the other hand the $b\bar b$ mesons are more compact and therefore respective chromoelectric polarizabilities are smaller. Very naively the polarizabilities in bottomonia are suppressed by the  factor $\left( \frac{\alpha_s(m_c) m_c}{\alpha_s(m_b) m_b} \right)^3$
(ratio of the Bohr radii cubed) in comparison with the polarizabilities in charmonia. With this naive estimate we obtain that the chromoelectric polarizability in bottomonia is close to the value that corresponds to the appearance of the nucleon - $\Upsilon(2S)$
bound state. It implies that more accurate calculations are required.  More detailed study of the interaction of higher excited quarkonia with the nucleon are also warranted.

\acknowledgments

This paper was supported by the NSF grant PHY-1402593. The work of V.~P. is supported by the Russian Science Foundation grant 14-22-00281. M.~V.~P. is grateful t    o Hyun-Chul Kim and Michal Praszalowicz for fruitful discussions.

%%%%%%%%%%%%%%%%%%%%%%%%%%%%%%%%%%%%%%%%%%%%%%%%%%%%%%

\end{document}